\documentclass[10pt,conference]{IEEEtran}
\makeatletter
\renewcommand{\@IEEEsectpunct}{~}
\makeatother

\ifCLASSINFOpdf
  \usepackage[pdftex]{graphicx}
\else
  \usepackage[dvips]{graphicx}
\fi
\usepackage{amsmath}

\usepackage{multicol, multirow, tabularx}
\usepackage[table]{xcolor}
\usepackage{url}
\usepackage{csquotes}
\usepackage[normalem]{ulem}
\usepackage{tcolorbox}
\usepackage{caption}

\usepackage{url}
\usepackage{csquotes}
\usepackage{threeparttable}

\begin{document}
%\title{What Challenges Do Developers Face About Managing Secrets-Containing Software Artifacts?}

\title{What Challenges Do Developers Face About Checked-in Secrets in Software Artifacts?}

\author{\IEEEauthorblockN{Setu Kumar Basak\IEEEauthorrefmark{1},
Lorenzo Neil\IEEEauthorrefmark{2}, Bradley Reaves\IEEEauthorrefmark{3} and
Laurie Williams\IEEEauthorrefmark{4}}
\IEEEauthorblockA{North Carolina State University, USA\\
Email: \IEEEauthorrefmark{1}sbasak4@ncsu.edu,
\IEEEauthorrefmark{2}lcneil@ncsu.edu,
\IEEEauthorrefmark{3}bgreaves@ncsu.edu,
\IEEEauthorrefmark{4}lawilli3@ncsu.edu}}

\maketitle

\begin{abstract}
Throughout 2021, GitGuardian's monitoring of public GitHub repositories revealed a two-fold increase in the number of secrets (database credentials, API keys, and other credentials) exposed compared to 2020, accumulating more than six million secrets. To our knowledge, the challenges developers face to avoid checked-in secrets are not yet characterized. \textit{The goal of our paper is to aid researchers and tool developers in understanding and prioritizing opportunities for future research and tool automation for mitigating checked-in secrets through an empirical investigation of challenges and solutions related to checked-in secrets}. We extract 779 questions related to checked-in secrets on Stack Exchange and apply qualitative analysis to determine the challenges and the solutions posed by others for each of the challenges. We identify 27 challenges and 13 solutions. The four most common challenges, in ranked order, are: (i) store/version of secrets during deployment; (ii) store/version of secrets in source code;  (iii) ignore/hide of secrets in source code; and (iv) sanitize VCS history. The three most common solutions, in ranked order, are: (i) move secrets out of source code/version control and use template config file; (ii) secret management in deployment; and (iii) use local environment variables. Our findings indicate that the same solution has been mentioned to mitigate multiple challenges. However, our findings also identify an increasing trend in questions lacking accepted solutions substantiating the need for future research and tool automation on managing secrets.

\end{abstract}


% \fi
\IEEEpeerreviewmaketitle

\section{Introduction} \label{Introduction}
%This demo file is intended to serve as a ``starter file''
%for IEEE conference papers produced under \LaTeX\ using
%IEEEtran.cls version 1.8b and later.
In March 2022, GitGuardian stated that the number of secrets exposed on public GitHub repositories doubled in 2021 compared to 2020, reaching a total of over six million secrets~\cite{GitGuardian}. To perform authentication across software artifacts as part of system integration, software developers need secrets (database credentials, API keys, and other credentials). During software development, these secrets may need to be shared by developers working on a team, and after deployment may need to be distributed to applications.

Version control system (VCS) repositories, such as GitHub~\cite{github} and GitLab~\cite{gitlab}, are widely used by developers for managing source code. However, the VCS repository's nature makes securing secrets in developer projects challenging. In 2019, Meli et al.~\cite{meli2019bad} studied a 13\% snapshot of public GitHub repositories and found over 200K API keys checked into the repositories. Secrets are not only pushed into VCS repositories by developers but also kept in Android and iOS application packages~\cite{android-leak}. Secrets in software artifacts (CWE-798: Use of Hard-coded Credentials~\cite{cwe-798}) have also been identified as a CWE Top 25 Most Dangerous Software Weaknesses~\cite{cwe-top-25}.

% Motivation
While the checked-in secrets issue is well-known through prior works~\cite{meli2019bad, 7180102, rahman2019share, rahman2019seven}, little is known about developers' technical challenges in preventing secrets from being stored in software artifacts. Developers query online forums, such as a developer who posted a question on how to keep secrets out of VCS repositories~\cite{stack-question-1}. Systematically analyzing questions asked by developers and solutions posed by others can reveal the technical challenges and practices adopted by the developers to protect the secrets.

\textit{The goal of our paper is to aid researchers and tool developers in understanding and prioritizing opportunities for future research and tool automation for mitigating checked-in secrets through an empirical investigation of challenges and solutions related to checked-in secrets.}

In this study, we analyze developers' questions and related solutions about checked-in secrets and provide answers to the following research questions:

\begin{itemize}
  \item \textbf{RQ1:} What are the technical challenges faced by developers related to checked-in secrets?
  \item \textbf{RQ2:} What solutions do developers get for mitigating checked-in secrets?
\end{itemize}

Users can post questions describing a particular technical challenge for which they need support on Stack Exchange~\cite{stackexchangesites}, a major question and answer (Q\&A) site. An answer is a suggestion or solution to a technical challenge. Users can pose multiple answers to a question, but either zero or one answer is accepted. The answer approved by the user who posted the question is termed as the \textit{accepted answer}. We refer to a question lacking an accepted answer or having no answers as a \textit{question with unsatisfactory answer}.

We extracted 779 questions related to checked-in secrets from Stack Exchange spanning from September 2008 to December 2021. From these questions, we conducted a qualitative analytical approach called card sorting~\cite{ZIMMERMANN2016137} to determine the question categories and related answer categories. We also perform quantitative analysis of question categories, which will help researchers and tool developers prioritize further study and tool development. In addition, the answer categories we presented give insights into which practices developers may have adopted. Following is a summary of the paper's contributions:
\begin{itemize}
  \item A set of challenges faced by the developers about checked-in secrets; and
  \item A set of solutions or suggestions posed by other developers to mitigate the checked-in secret challenges
\end{itemize}

The rest of our paper is structured as follows: The methodology used in our work is described in Section \ref{Methodology}. We discuss our findings and recommendations in Section \ref{Results} and \ref{Discussion}, respectively. The ethics and limitations of our paper is discussed  in Section \ref{Ethics} and \ref{ThreatToValidity}, respectively. Section \ref{RelatedWork} summarizes previous research findings pertinent to our paper. Finally, Section \ref{Conclusion} draws the paper's conclusion.

\section{Methodology} \label{Methodology}
We provide our four-step process for data collection and question and answer analysis as follows:

\subsection{Step 1: Q\&A Site Selection}

For collecting questions related to checked-in secrets, we selected Stack Exchange~\cite{stackexchangesites} which has been extensively used to gain insights from developers' questions to align future research and guide tools providers~\cite{TAHIR2020106333, rahman2018questions}. Stack Exchange consists of 179 Q\&A sites~\cite{stackexchangesites}. We extract the name and description of all the sites and manually read them. Then, we select sites that allow questions related to software development, software engineering, and software security. For example, the site ``Software Engineering'' can feature queries from developers, according to the site description ``Q\&A for professionals, academics, and students working within the systems development life cycle''.
The first author selected three Q\&A sites: ``Stack Overflow''~\cite{stackoverflow}, ``Information Security''~\cite{informationsecurity} and ``Software Engineering''~\cite{softwareengineering}. The basic statistics of the three sites are shown in Table~\ref{sites-stats}. In Step 2, we use these sites for question collection.

\begin{table} [!htb]
\begin{center}
\caption[abc]{Basic statistics of Stack Overflow (SO), Information Security (IS) and Software Engineering (SE) sites\footnotemark}
\label{sites-stats}
\footnotesize
\begin{tabular}{|c| c| c|c|c|} 
 \hline
 \textbf{Site} & \textbf{\#Questions} & \textbf{\#Answers} & \textbf{\#Users} & \textbf{\#Questions/Day} \\ [0.5ex] 
 \hline\hline
 SO & 23m & 34m & 18m & 5.5k \\ 
 \hline
 IS & 66k & 114k & 228k & 9.6 \\
 \hline
 SE & 61k & 173k & 352k & 5.5 \\
 \hline
\end{tabular}
\end{center}
\end{table}

\footnotetext{Based on data retrieved from the Stack Exchange Data Explorer \cite{stackexchangesites} on June 2022}

\subsection{Step 2: Content Collection}

\textbf{Start with initial tags and keywords for title and body:} To increase the likelihood of speedy response and aid in automated search, each question can be given one or more tags~\cite{stackexchangeexp}. Tags allow the extraction of questions that are specific to a given technology. For example, the tag ``secret-key'' can be used for identifying questions related to checked-in secrets according to the tag description ``Use this tag for questions related to the creation, storage and usage of secret keys''. Initially, we select ``secret-key'' and ``access-keys'' tags. Users can also post questions without giving tags. To avoid missing candidate questions, we use secrets-related keywords, such as ``expose'', ``protect'', and ``sensitive'', to search in the body and title of the questions to extract relevant questions. 

\textbf{Extract questions from Stack Exchange data explorer:}
The Stack Exchange dataset is accessible publicly via data dumps~\cite{stackexchangedump} and the Stack Exchange data explorer~\cite{stackexchangeexp}. The data dumps are released quarterly, whereas the online Stack Exchange data explorer provides the most recent data. We use the tags and keywords in a SQL query and extract data from the Stack Exchange data explorer instead of data dumps. We collect the ID, title, body, accepted answer, view count, score, creation date, closed date, and tags of each extracted question from the three sites identified in Step 1. We collected 6022, 2591, and 1415 questions from Stack Overflow, Information Security, and Software Engineering sites, respectively.

\textbf{Identify relevant questions:}
We manually inspected each question's title and body and accepted questions with a  discussion related to checked-in secrets while rejecting all others.

\textbf{Find new relevant tags and keywords:}
We use snowball sampling~\cite{snowballsampling} which is a non-probability sample selection technique to locate hidden populations by relying on the characteristics of initial sample. Since a question can have multiple tags, we find new relevant tags by looking at all the tags present in the questions. For example, the question ``Where to keep static information securely in Android app?''~\cite{so-ex-tag} can be found by ``secret-key'' tag. The question also has tags ``access-token'' and ``security'' which we can add to our list of tags for finding more questions. Similarly, add new keywords by reading the title and body of the question. Altogether, we used 59 tags and 42 keywords which can be found in Table~\ref{tags-keywords-table}.

\newcolumntype{b}{>{\hsize=0.670\hsize}X}
\newcolumntype{s}{>{\hsize=0.330\hsize}X}
\begin{table} [!htb]
%\small
\footnotesize
\caption{List of Tags and Keywords used to extract questions from Stack Exchange sites}
\label{tags-keywords-table}
\begin{tabularx}{\columnwidth} {|b | s |}
 \hline
 \multicolumn{1}{|c|}{\textbf{Tags}} &
  \multicolumn{1}{c|}{\textbf{Keywords}}\\
 \hline \hline
 secret-key, access-keys, access-token, security, credentials, passwords, api-key, private-key, app-secret, connection-string, sensitive-data, environment-variables, config-files, certificate, configuration, google-api, amazon-s3, oauth, youtube-api, stripe-api, square, paypal, braintree, amazon-mws, gmail-api, twilio-api, mailgun, mailchimp, google-drive-api, key-management, development-process, coding-style, password-protection, source-code-protection, code-security, source-code, secure-coding, open-source, azure-key-vault, password-storage, password-management, key-exchange, confidentiality, sensitive-data-exposure, web-development, git, gitignore, version-control, github, svn, tfs, gitlab, repository, bitbucket, launchpad, mercurial, git-rewrite-history, git-history, git-filter-branch  
 
 & expose, exposing, protect, protecting, sensitive, remove, removing, commit, committing, share, sharing, keep, keeping, manage, managing, delete, deleting, clear, clearing, ignore, ignoring, secure, securing, store, storing, hide, hiding, avoid, avoiding, push, pushing, host, hosting, security, connection string, secret, password, credential, private key, token, api key, access key  \\
\hline
\end{tabularx}
\end{table}

\textbf{Repeat and stop criteria:}
We repeat the previous step until we no longer found new tags and keywords in each set of extracted questions.

Finally, we identified 694 questions in Stack Overflow, 40 questions in Information Security, and 45 questions in Software Engineering. In total, we identified 779 questions from the three sites spanning from September 2008 to December 2021 which are available online~\cite{repo-artifacts}. The count of questions from each year before and after filtering is shown in Table~\ref{question-count}.

\begin{table} [!htb]
\begin{center}
\begin{threeparttable}
\caption{Question Count Per Year for Stack Overflow (SO), Information Security (IS) and Software Engineering (SE) sites}
\label{question-count}
%\small
\footnotesize
\begin{tabular}{|c| c| c| c|c|c|c|c|c|} 
 \hline
 \textbf{Year} & \textbf{SO}\tnote{a} & \textbf{SO}\tnote{b} & \textbf{IS}\tnote{a} & \textbf{IS}\tnote{b} & \textbf{SE}\tnote{a} & \textbf{SE}\tnote{b} & \textbf{Total}\tnote{a} & \textbf{Total}\tnote{b}\\ 
 \hline\hline
 2008 & 23 & 4 & 0 & 0 & 0 & 0 & 23 & 4 \\ \hline
 2009 & 136 & 22 & 0 & 0 & 0 & 0 & 136 & 22 \\ \hline
 2010 & 212 & 30 & 5 & 0 & 28 & 1 & 245 & 31 \\ \hline
 2011 & 284 & 43 & 73 & 0 & 163 & 5 & 520 & 48 \\ \hline
 2012 & 370 & 44 & 129 & 2 & 170 & 8 & 669 & 54 \\ \hline
 2013 & 447 & 48 & 160 & 6 & 146 & 7 & 753 & 61 \\ \hline
 2014 & 485 & 41 & 257 & 5 & 136 & 3 & 878 & 49 \\ \hline
 2015 & 481 & 47 & 361 & 5 & 152 & 4 & 994 & 56 \\ \hline
 2016 & 581 & 51 & 340 & 6 & 126 & 2 & 1047 & 59 \\ \hline
 2017 & 581 & 76 & 323 & 5 & 110 & 4 & 1014 & 85 \\ \hline
 2018 & 538 & 63 & 268 & 5 & 107 & 4 & 913 & 72 \\ \hline
 2019 & 518 & 54 & 235 & 1 & 102 & 3 & 855 & 58 \\ \hline
 2020 & 722 & 88 & 226 & 1 & 90 & 2 & 1038 & 51 \\ \hline
 2021 & 644 & 83 & 214 & 4 & 85 & 2 & 943 & 89 \\ \hline
\end{tabular}
\begin{tablenotes}\footnotesize
\item [a] Total number of questions before filtering
\item [b] Total number of questions after filtering
\end{tablenotes}
\end{threeparttable}
\end{center}
\end{table}

\subsection{Step 3: Identifying Question and Answer Categories}
From the 779 checked-in secrets-related questions, two authors independently apply card sorting~\cite{ZIMMERMANN2016137}, a qualitative analysis technique, to identify the question and answer categories. Card sorting is a qualitative technique for classifying textual items into categories~\cite{ZIMMERMANN2016137}. Card sorting aids in creating informative categories and is commonly used in research~\cite{rahman2018questions}. The following three phases of card sorting are implemented in accordance with Zimmerman et al.~\cite{ZIMMERMANN2016137}'s recommendations.

\textbf{\uline{Preparation:}} Each question's ID, title, body, and accepted answer are collected.

\textbf{\uline{Execution:}}
The first and second authors perform card sorting by giving labels to each question and the corresponding answer and sort into categories. The body and title of the questions are used to derive question categories, whereas the accepted answers are used to derive answer categories.

\textbf{\uline{Analysis:}}
The obtained question and answer categories are cross-checked by both authors after the first and second authors finish their card sorting analysis individually. We use a negotiated agreement~\cite{negotiatedagreement} to resolve the disagreed-upon categories. A negotiated agreement is an approach to discuss the disagreements among the raters to resolve disagreements when two or more raters code the same artifacts~\cite{negotiatedagreement}. We resolve disagreements by discarding categories inappropriate for checked-in secrets or combining similar categories into one category. The first author determines 32 unique question categories and 16 unique answer categories. The second author determines 30 unique question categories and 14 unique answer categories. The first and second authors finalize 27 question and 13 answer categories by resolving the disagreements presented in Table \ref{question-desc-example} and Section \ref{Results}, respectively.

\subsection{Step 4: Analysis}

We use the identified question and answer categories from Step 3 to answer our research questions. 

\subsubsection{RQ1: What are the technical challenges faced by developers related to checked-in secrets?}

We break down RQ1 into four sub-research questions as below:

\begin{itemize}
    \item \textbf{RQ1.1} What are the questions developers ask about checked-in secrets? 
    \item \textbf{RQ1.2} Which questions related to checked-in secrets exhibit more unsatisfactory answers?
    \item \textbf{RQ1.3} Which questions are the most popular among developers related to checked-in secrets? 
    \item \textbf{RQ1.4} How do question categories related to checked-in secrets trend over time?
\end{itemize}    

We investigate the four sub-research questions as following:

\textbf{RQ1.1: What are the questions developers ask about checked-in secrets?} We first provide the set of question categories to answer RQ1.1 along with a description and an example of each category which we determine from Step 3. Next, we compute the proportion of questions for each category $x$, QC($x$).

\textbf{RQ1.2: Which questions related to checked-in secrets exhibit more unsatisfactory answers?} A question with no accepted answer could indicate that the developer who asked the question was dissatisfied with the responses. Lacking accepted answers or having no answers may suggest an important category that needs assistance. We answer RQ1.2 by quantifying which of the checked-in secrets-related question categories has more questions with unsatisfactory answers. We compute the proportion of questions with unsatisfactory answers for question category $x$, UNC($x$). 

Furthermore, we compute the proportion of questions with unsatisfactory answers for each year $y$, TUN($y$) to see how the proportion of unsatisfactory answers related to checked-in secrets has changed over time.

\textbf{RQ1.3: Which questions are the most popular among developers related to checked-in secrets?} Developers can view a question and corresponding answers without becoming registered users on Stack Exchange. The number of total visits for a question by registered and non-registered users of the website is used to calculate the View Count of a question~\cite{stackexchangeexp}. The View Count can help us observe which questions are most popular among the developers. Registered users can also vote up or down on questions. Upvotes indicate that users find the question helpful, well-researched, or thought-provoking. Downvotes indicate that users believe the question lacks real explanation, contains misleading information, or is poorly researched. A question's Score on Stack Exchange is calculated by subtracting the number of downvotes from the number of upvotes~\cite{yao2013want}. Rather than selecting a single metric, we use both View Count and the Score of the question as a better approximation for question popularity. Previous studies use a similar a popularity metric~\cite{TAHIR2020106333}.

We use Spearman's rho $\rho$~\cite{spearmanrho} to verify the rank correlation between View Count and Score. View Count is found to have a significant correlation with Score ($\rho$ = 0.72, $\alpha < $ 0.001). We use Feature Scaling \cite{featurescale} to normalize the View Count and Score values of each question by Equation \ref{eqnorm} since the range of both the metrics are different. 

\begin{equation} \label{eqnorm}
        X_{nor} = \frac{X - X_{min}}{X_{max} - X_{min}}
    \end{equation}

where $X$ denotes the original value, $X_{min}$ denotes the range's minimum value, $X_{max}$ denotes the range's maximum value and $X_{nor}$ denotes the normalized value.    

To determine how popular a question is, we use the average of normalized View Count and Score values. Next, we calculate the popularity of each category $x$, PQ($x$) using Equation \ref{eqnpopcat}. A question category $x$ with a high popularity score means developers need support to mitigate the specific challenge. \vspace{-1em} 

\begin{gather} \label{eqnpopcat}
    \begin{aligned}[b]
        \textnormal{PQ(\(x\))} =
        \frac{\textnormal{sum of popularity score of questions in category \(x\)}}{\textnormal{total questions in category \(x\)}}
   \end{aligned}  
\end{gather}

\textbf{RQ1.4: How do question categories related to checked-in secrets trend over time?} We examine temporal trends, similar to previous studies~\cite{rahman2018questions, bajaj2014mining}, to see how the number of questions relevant to the identified question categories changes over time. We first use Equation \ref{eqtrend} to compute the temporal trend of category $x$ for each month $m$. 

\begin{gather} \label{eqtrend}
    \begin{aligned}[b]
        \textnormal{TT(\(x\), \(m\))} = 
        \frac{\textnormal{number of questions of category \(x\) in month \(m\)}}{\textnormal{number of questions in month \(m\)}}
   \end{aligned}  
\end{gather}

Then, to see whether the observed trend is significantly increasing or decreasing, we use the Cox-Stuart test~\cite{coxstuarttest}, a statistical method that compares earlier data points in a time series to later data points to evaluate the trend. To assess which question categories have increasing or decreasing trends, we apply a 95\% statistical confidence level ($p<0.05$). We term the temporal trend to be ``Consistent'' if we can not determine whether the trend is increasing or decreasing.

\subsubsection{RQ2: What solutions do developers get for mitigating checked-in secrets?}

To answer RQ2, we first provide the answer categories to mitigate the challenges related to checked-in secrets, which we determine from Step 3. Then, we provide a mapping of answer categories to each of the question categories. From the question-answer category mapping, we can understand the solutions posed by developers to mitigate a specific technical challenge.

\section{Results} \label{Results}
In this section, we discuss our findings and answer our research questions.

\subsection{Answer to RQ1: What are the technical challenges faced by developers related to checked-in secrets?}
We answer the four sub-research questions of RQ1 in the following sub sections.

\subsubsection{\uline{Answer to RQ1.1: What are the questions developers ask about checked-in secrets?}}

We identify 27 unique question categories of 9 domains, which we present in Table \ref{question-desc-example} sorted based on the number of questions in a domain. The domain name, question category name, a description of the question category, and a representative example are provided for all the question categories. The number of questions in each category is indicated in parenthesis in the ``Category'' column.

The proportion of questions in each identified question category and the other four metrics mentioned in Section \ref{Methodology} are presented in Table \ref{question-count-pct}. The proportion of questions, percentage of unsatisfactory answers, popularity score, Cox-Stuart test value of temporal trend of questions, and the identified trend of questions in each question category are represented in the columns ``QC(\%)'', ``UNC(\%) (Count)'', ``PQ'', ``Cox Stuart, p-value'' and ``Trend'' respectively. According to Table \ref{question-count-pct}, the top four question categories based on QC metric are ``(Deployment) Store/Version'', ``(Secrets) Store/Version'', ``(Secrets) Ignore/Hide'', and ``(VCS Feature) History Sanitize''. These four categories constitute 56.1\% of all questions.

  \newcolumntype{a}{>{\hsize=0.03\hsize}X}
\newcolumntype{b}{>{\hsize=0.745\hsize}X}
\newcolumntype{m}{>{\hsize=0.14\hsize}X}
\newcolumntype{s}{>{\hsize=0.14\hsize}X}
\begin{table*}
\centering
\caption{27 question categories. References to all the examples and developer quotes are available online \cite{figshare-links}}
\label{question-desc-example}
%\scriptsize
\footnotesize
%\tiny
\begin{tabularx}{\textwidth} {|a|m|b|s|}
\hline
\multicolumn{1}{|c|}{\textbf{Domain}} & \multicolumn{1}{l|}{\textbf{Category (Count)}} & \textbf{Description} & \textbf{Example} \\ \hline \hline
\multirow{5}{*}{Secrets}
 & Q1: Store/Version (121) & We observe that the same questions of knowing the best way to store secrets have been asked for different technologies, such as ASP.NET and Python. We also observe developers asking about versioning the secrets for environments, such as development and production environments, where they do not know the consequences of storing secrets in VCS repositories. & \textit{How should I store a password used by a service written in .NET?} \\ \cline{2-4} 
 
 & Q2: Ignore/Hide (85) & We observe that developers are aware of the consequences of secrets presence in the source code and want to hide the secrets. As one developer stated: \textit{``The credentials are hard-coded at the moment, but they should not be. What is the proper way of hiding them?''}. Developers also question about challenges faced in avoiding secrets from being committed to the VCS repository.        & \textit{Hide API keys from github public repo?} \\ \cline{2-4}

 & Q3: Exploitability (30) & Developers do not know whether storing a secret such as a Google API key or testing credentials in source code or a VCS repository can be exploited. For example, one developer stated: \textit{``I'm making use of google API for location. Can the key be hardcoded? ... If it's sensitive, why is it sensitive and how can attackers exploit this?''}. & \textit{Is having sensitive data in a PHP script secure?} \\ \cline{2-4} 
 
  & Q4: Distribute (11) & Developers ask questions about sharing secrets with other developers so that they can run the project successfully in their environment. As one developer stated: \textit{``How can I keep my API key secret, but have my project still be functional if someone clones the repo?''}. We observe that developers are unsure how to share secrets with specific developers without exposing them.     & \textit{Push to GitHub that project is still functional when the repo cloned?} \\ \cline{2-4} 
 
 & Q5: Restriction (2)  & We observe questions posted for restricting a specific group of developers from having access to secrets. For example, \textit{``What happens if a malicious developer decides to steal the secret (say, an API key) and use it for malicious purposes? Is there a way to store secrets such that a backend developer doesn't have direct access to the API Key?''}. & \textit{What are ways to manage secrets in a big organisation?} \\ \hline
 
 \multirow{3}{*}{\begin{tabular}[l]{@{}l@{}}Deploy-\\ment\end{tabular}} 
 & Q6: Store/Version (149) & Platform as a service (PaaS), such as Heroku \cite{heroku} and Google App Engine \cite{google-app-engine}, are commonly used to manage applications. During deployment, the code is fetched from the repositories. We observe developers asking questions about where to store the secrets needed for deployment since secrets are not pushed in the repository. Developers want to know the secure way of versioning secrets for deployment environments. This question category is the most frequently asked. & \textit{Where to store sensitive files for heroku platform?} \\ \cline{2-4}
 
  & Q7: Improper Configuration (34) & As the configuration (config) files are ignored in the repository and source code is fetched from the repository for deployment, developers are getting exceptions due to improper configuration in the deployment server. We observe developers asking for help resolving the build and deploy-related exceptions. We observe that most of the exceptions are during Django application deployment. & \textit{Azure Django App has SECRET\_KEY Exception} \\ \cline{2-4} 
 
 & Q8: Ignore/Hide (15) & During the build and deployment of an application, developers use the secrets present in the continuous integration and continuous deployment (CI/CD) scripts or the VCS repository. We observe developers asking to know the best practice of hiding the secrets from CI/CD scripts or repositories and perform successful build and deploy.  & \textit{Docker-Compose with Gitlab CI managing sensitive data} \\ \cline{2-4} 
 
 & Q9: Dot File (3)  & Developers deploy directly from VCS repositories using Git tools. They push sensitive dot files such as .git and .gitignore files that can be accessed at the website's root location. Previous research \cite{meli2019bad} has found secrets in the .gitignore file, even though the .gitignore file is designed to restrict unintended source files committing into VCS. We observe developers facing challenges restricting the dot files' access from the website's root. & \textit{How to make .gitignore safe?} \\ \hline
 
 \multirow{4}{*} {\begin{tabular}[l]{@{}l@{}}VCS \\Feature\end{tabular}} 
  & Q10: History Sanitize (81) & Developers accidentally or knowingly push sensitive information into the VCS repository. One developer stated: \textit{``I am using a shared github repository to collaborate on a project ... I committed and pushed a script file containing a password which I don't want to share''}. The sensitive information remains in the VCS history even when removed in another commit. Developers ask questions about sanitizing the VCS history using different tools but could not use the tools properly. Rahman et al. \cite{rahman2022secret} also observed developers bypassing secret scanning tools warning because of facing technical challenges of eliminating secrets completely from the VCS history. & \textit{How to remove sensitive data from a file in github history?} \\ \cline{2-4} 
  
    & Q11: Ignore Already Committed (14)  & Knowing the exploitability of secrets present in source code, developers want to commit a default file without secrets. However, they want to untrack further local changes of the file from VCS repositories to avoid accidentally committing the local changes, and VCS does not support the functionality \cite{ignore-already-committed}. As a result, we observe developers ask questions about ignoring an already-committed file from VCS tracking. & \textit{Stop tracking file in Git after a first commit?} \\ \cline{2-4}
  
  & Q12: Line Level Security (11) & \textit{``Do any version control systems allow you to specify line level security restrictions rather than file level?''} stated by one developer. VCS, such as Git, only supports file-level restrictions. We observe developers wanting to mark specific lines in a file that contains secrets and tell the VCS to secure the lines to avoid exposing the secrets. & \textit{hide or change value a line at git commit but not locally} \\ \cline{2-4}
  
   & Q13: Encrypt File (1) & We observe developers asking questions about if there is a way to encrypt a secrets-containing file before committing to VCS repositories. & \textit{Encrypting files added to repos} \\
   \hline
 
 \multirow{5}{*} {\begin{tabular}[l]{@{}l@{}}Configur-\\ ation File\end{tabular}}   
 & Q14: Store/Version (56) & Config files contain secrets. We observe developers face challenges storing the config files in the VCS repository since it would expose the secrets. For example, one developer stated: \textit{``I'd like to version control the whole project, including config file, but I don't want to share my passwords''}. & \textit{Preferred way to store application configurations?} \\ \cline{2-4} 
 
  & Q15: Ignore/Hide (32) & We observe developers asking questions about ignoring or hiding sensitive secrets-containing config files such as the web.config and database.yml files from the VCS repository. Developers also complain about the lack of documentation or suggestions the specific technology provides on ignoring config files. & \textit{Protecting the sensitive files from pushing to version control?} \\ \cline{2-4} 
  
    & Q16: Distribute (9) & Developers face challenges sharing secrets-containing config files with other team members without exposing them publicly. For example, one developer stated: \textit{``Should I add these 2 files to versioning or do I have to distribute these files manually to other team members?''}. & \textit{Managing project config files in repository?} \\ \cline{2-4} 
    
    & Q17: Exploitability (3) & Developers place environment variables replacing secrets in the config files and want to confirm the exploitability from outside. We also observe developers placing secrets in PHP .ini files and asking about the exploitability of the secrets. For example, one developer stated: \textit{``Is better to hide somewhere .ini file and deny access via .htaccess?''}. & \textit{Storing sensitive info. inside .ini file is good or bad approach?} \\ \cline{2-4}
    
     & Q18: Accessibility (3)  & To avoid exposing secrets, developers load secrets dynamically by referencing external files in config files but get an undefined error. An example includes loading an external database settings file into a web.config file. We observe developers facing challenges in avoiding the undefined error and could not find the proper documentation. & \textit{How to securely use credentials outside web.config?} \\ \hline
     
     \multirow{1}{*}{\begin{tabular}[l]{@{}l@{}}Pre-open\\ Source\end{tabular}} 
 & Q19: Cross-check (52) & We observe developers asking questions before open-sourcing their projects. The questions include should developers clean VCS history and what checklists should they run to avoid exposing secrets. & \textit{OpenAuth \& Open Source Projects?} \\

 \hline
\end{tabularx}
\end{table*}
  \newcolumntype{a}{>{\hsize=0.03\hsize}X}
\newcolumntype{b}{>{\hsize=0.745\hsize}X}
\newcolumntype{m}{>{\hsize=0.14\hsize}X}
\newcolumntype{s}{>{\hsize=0.14\hsize}X}
\begin{table*}
\ContinuedFloat
\centering
\caption{27 question categories (Continued). References to all the examples and developer quotes are available online \cite{figshare-links}}
\label{question-desc-example}
%\scriptsize
\footnotesize
%\tiny
\begin{tabularx}{\textwidth} {|a|m|b|s|}
\hline
\multicolumn{1}{|c|}{\textbf{Domain}} & \multicolumn{1}{l|}{\textbf{Category (Count)}} & \textbf{Description} & \textbf{Example} \\ \hline \hline
  
   \multirow{3}{*} {\begin{tabular}[l]{@{}l@{}}Client- \\Side\\Applicat-\\ion\end{tabular}} 
  & Q20: Store (28) & Developers work on client-side applications without a server-side implementation and store secrets on the client-side, such as in Javascript and Android applications. Developers face challenges in storing the secrets securely as secrets can easily be exposed from the developer console or by decompiling the binary packages. & \textit{Securely storing secret data in a client-side web application?} \\ \cline{2-4} 
  
  & Q21: Hide (14) & One developer stated: \textit{``Using Javascript however, I don't feel comfortable that the client secret is exposed in my code ... because if someone looks at my source they have the client\_id and client\_secret which makes it possible to authenticate themselves with my code''}. We observe developers looking for ways to hide client-side application secrets. & \textit{How do I hide API key in create-react-app?} \\ \cline{2-4} 
  
     & Q22: Exploitability (5)  & Developers ask questions to confirm whether the implementation of keeping secrets in the client-side application code is exploitable or not. \textit{``Could I sleep at night knowing that I won't see ``Super Cool Web App Hacked, change your passwords!'' all over HN and Reddit ... as a result of this implementation.''} stated by one developer. & \textit{In iOS, is there leak risk if I write the secret key in the code?} \\ \hline
     
         \multirow{2}{*}{\begin{tabular}[l]{@{}l@{}}Secure-\\ness\end{tabular}} 
 & Q23: Private Repository (13) & One developer stated: \textit{``Is it safe for me to store my Amazon S3 keys/secrets in a private Github repo? I know that it is not safe for a public repo but I am wondering if a private repo is safe?''}. We observe developers asking about the safety of secrets present in a private repository. & \textit{Storing Amazon S3 keys in private repo} \\ \cline{2-4}
 
  & Q24: Unpushed Branch (1) & We observe developers ask questions about the security of secrets if they do not push the secrets-containing branch to a public repository. For example, one developer stated: ``Is there any chance my sensitive data could end up in the remote repository index somehow?''. & \textit{Commit password to branch that never pushed?}\\ 
  \hline

  \multirow{1}{*}{\begin{tabular}[l]{@{}l@{}}External\\ Secret \\Manage-\\ment\end{tabular}} 
 & Q25: Setup (3) & We observe developers moving secrets to external secret management services, such as HashiCorp Vault \cite{hashicorp-vault} and Azure Key Vault \cite{azure-vault}. However, developers face challenges in properly setting up these hardware security modules. Examples of such questions include where to store the vault key, the feasibility of using vaults, and how to store the database connection strings in the vault. & \textit{Storing DB Connection Strings in Azure Key Vault} \\ \hline
  
  \multirow{2}{*}{Others} 
 & Q26: Importance (2) & We observe developers asking questions about why they should keep secrets out of the VCS repository. For example, one developer stated: \textit{``It seems like common knowledge that it's a good practice to keep secrets files ... checked out of your git repository ... Why?''}. & \textit{Why should you keep secrets out of your repository?} \\ \cline{2-4}
 
  & Q27: Decision (1) & One developer stated: \textit{``Today I found what looked to be my supervisor's password in some code in version control ... How should I handle this situation?''}. We observe developers being hesitant about making decisions when they find secrets in the VCS repository. & \textit{What should I do when I find sensitive info in VCS?}\\ 
  \hline
     
\end{tabularx}
\end{table*}

\subsubsection{\uline{Answer to RQ1.2: Which questions related to checked-in secrets exhibit more unsatisfactory answers?}}
%based on the Q metric
Table \ref{question-count-pct} shows that UNC scores of more than 40\% are found in 16 of the 27 identified question categories. Our finding indicates that 44.3\% of questions within our dataset have unsatisfactory answers. The top four question categories, ``(Deployment) Store/Version'', ``(Secrets) Store/Version'', ``(Secrets) Ignore/Hide'' and ``(VCS Feature) History Sanitize'' have UNC scores of 43.0\%, 47.1\%, 34.1\% and 45.7\% respectively.

\begin{figure}
    \includegraphics[width=\columnwidth]{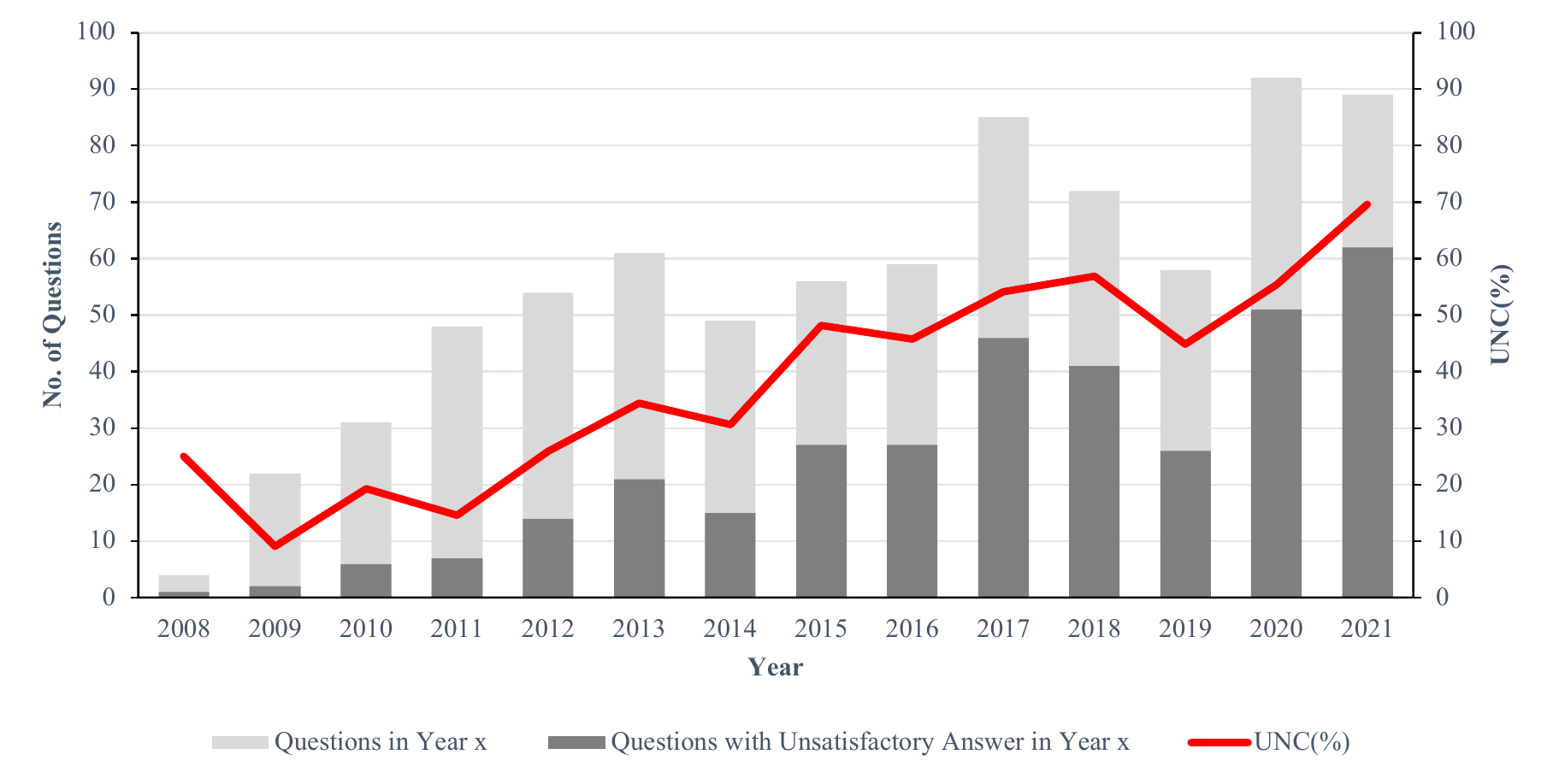}
    \caption{Trend of Unsatisfactory Answer Per Year}
    \label{fig:answer_trend_per_year}  
\end{figure}

Figure \ref{fig:answer_trend_per_year} presents the trend of unsatisfactory answers for each year between 2008 and 2021. We observe that the percentage of unsatisfactory answers shows an increasing trend. More than 50\% of questions have unsatisfactory answers since 2017, thus indicating that the developers are not getting desired answers to mitigate the challenges of checked-in secrets.

\newcolumntype{b}{>{\hsize=0.480\hsize}X < {\centering}}
\newcolumntype{s}{>{\hsize=0.125\hsize}{c} < {\centering}}
\begin{table*}[!t]
\centering
\caption{Summary of identified question categories, sorted by decreasing question proportion (QC)}
\label{question-count-pct}
%\scriptsize
\footnotesize
%\small
\begin{tabularx}{0.830\textwidth} {|l | c | c| c | c| c|}
 \hline
 \multicolumn{1}{|l|}{\textbf{(Domain) Question Category}} &
  \multicolumn{1}{c|}{\textbf{QC (\%)}} &
  \multicolumn{1}{c|}{\textbf{UNC (\%) (Count)}} &
  \multicolumn{1}{c|}{\textbf{PQ}} &
  \multicolumn{1}{c|}{\textbf{Cox Stuart, p-value}} &
  \multicolumn{1}{c|}{\textbf{Trend}} \\
 \hline \hline
 (Deployment) Store/Version & 19.2 & 43.0 (64) & 0.020 & $\Uparrow$, 0.11 & Consistent  \\ \hline
 (Secrets) Store/Version & 15.6  & 47.1 (57) & 0.023 & $\Uparrow$, 0.003 & \cellcolor[gray]{0.8} Increasing \\ \hline
 (Secrets) Ignore/Hide & 10.9  & 34.1 (29) & 0.015 & $\Uparrow$, 0.11 & Consistent \\ \hline
 (VCS Feature) History Sanitize & 10.4 & 45.7 (37) & 0.018 & $\Uparrow$, $<$ 0.001 & \cellcolor[gray]{0.8} Increasing \\ \hline
 (Configuration File) Store/Version & 7.2 & 39.3 (22) & 0.022 & $\Uparrow$, 0.5 & Consistent \\ \hline
 (Pre-open Source) Cross-check & 6.7 & 40.4 (21) & 0.010 & $\Downarrow$, 0.3 & Consistent \\ \hline
 (Deployment) Improper Configuration & 4.4 & 58.8 (20) & 0.008 & $\Uparrow$, $<$ 0.001 & \cellcolor[gray]{0.8} Increasing \\ \hline
 (Configuration File) Ignore/Hide & 4.1 & 40.6 (13) & 0.008 & $\Downarrow$, 0.34 & Consistent \\ \hline
 (Secrets) Exploitability & 3.9 & 56.7 (17) & 0.014 & $\Downarrow$, 0.59 & Consistent \\ \hline
 (Client-Side Application) Store & 3.6 & 60.7 (17) & 0.030 & $\Uparrow$, 0.002 & \cellcolor[gray]{0.8} Increasing \\ \hline
 (Deployment) Ignore/Hide & 1.9 & 46.7 (7) & 0.010 & $\Uparrow$, 0.09 & Consistent \\ \hline
 (VCS Feature) Ignore Already Committed & 1.8 & 28.6 (4) & 0.007 & $\Uparrow$, 0.29 & Consistent \\ \hline
 (Client-Side Application) Hide & 1.8 & 35.7 (5) & 0.022 & $\Uparrow$, 0.13 & Consistent \\ \hline
 (Secureness) Private Repository & 1.7 & 46.2 (6) & 0.014 & $\Uparrow$, 0.27 & Consistent \\ \hline
 (Secrets) Distribute & 1.4 & 63.6 (7) & 0.007 & $\Uparrow$, 0.11 & Consistent \\ \hline
 (VCS Feature) Line Level Security & 1.4 & 36.4 (4) & 0.007 & $\Downarrow$, 0.5 & Consistent \\ \hline
 (Configuration File) Distribute & 1.2 & 66.7 (6) & 0.007 & $\Uparrow$, 0.14 & Consistent \\ \hline
 (Client-Side Application) Exploitability & 0.6 & 40.0 (2) & 0.008 & $\Uparrow$, 0.5 & Consistent \\ \hline
 (Configuration File) Exploitability & 0.4 & 0.0 (0) & 0.012 & $\Uparrow$, 0.5 & Consistent \\ \hline
 (Configuration File) Accessibility & 0.4 & 33.3 (1) & 0.008 & $\Uparrow$, 0.13 & Consistent \\ \hline
 (Deployment) Dot File & 0.4 & 33.3 (1) & 0.007 & $\Uparrow$, 0.5 & Consistent \\ \hline
 (External Secret Management) Setup & 0.4 & 66.7 (2) & 0.015 & $\Uparrow$, 0.13 & Consistent \\ \hline
 (Others) Importance & 0.3 & 100.0 (2) & 0.005 & $\Uparrow$, 0.25 & Consistent \\ \hline
 (Secrets) Restriction & 0.3 & 50.0 (1) & 0.007 & $\Downarrow$, 0.75 & Consistent \\ \hline
 (VCS Feature) Encrypt File & 0.1 & 0.0 (0) & 0.008 & $\Downarrow$, 0.5 & Consistent \\ \hline
 (Secureness) Unpushed Branch & 0.1 & 0.0 (0) & 0.005 & $\Downarrow$, 0.5 & Consistent \\ \hline
 (Others) Decision & 0.1 & 0.0 (0) & 0.008 & $\Downarrow$, 0.5 & Consistent \\ \hline
\end{tabularx}
\end{table*}
\newcolumntype{b}{>{\hsize=0.95\hsize}X}
\newcolumntype{s}{>{\hsize=0.05\hsize}X}
\begin{table*}
\caption{Ranked Order of Question Categories Based on Popularity (PQ) and Unsatisfactory Answer Percentage (UNC)}
\label{question-cat-hierarchy}
%\normalsize
\footnotesize
%\small
\begin{tabularx}{\textwidth} {|s | b |}
 \hline
 \multicolumn{1}{|l|}{\textbf{Metric}} &
  \multicolumn{1}{c|}{\textbf{(Domain) Question Category (Sorted in decreasing order of metric)}}\\
 \hline \hline
 PQ &
 (Client-Side Application) Store,
 (Secrets) Store/Version, 
 (Client-Side Application) Hide,
 (Configuration File) Store/Version,
 (Deployment) Store/Version,
 (VCS Feature) History Sanitize,
 (Secrets) Ignore/Hide,
 (External Secret Management) Setup,
 (Secureness) Private Repository,
 (Secrets) Exploitability,
 (Configuration File) Exploitability,
 (Pre-open Source) Cross-check,
 (Deployment) Ignore/Hide,
 (VCS Feature) Encrypt File,
 (Configuration File) Accessibility,
 (Others) Decision,
 (Configuration File) Ignore/Hide,
 (Client-Side Application) Exploitability,
 (Deployment) Improper Configuration,
 (Deployment) Dot File,
 (Secrets) Restriction,
 (Secrets) Distribute,
 (Configuration File) Distribute,
 (VCS Feature) Line Level Security,
 (VCS Feature) Ignore Already Committed,
 (Others) Importance,
 (Secureness) Unpushed Branch\\
 \hline
 UNC (\%) &
 (Others) Importance,
 (Configuration File) Distribute,
 (External Secret Management) Setup,
 (Secrets) Distribute,
 (Client-Side Application) Store,
 (Deployment) Improper Configuration,
 (Secrets) Exploitability,
 (Secrets) Restriction,
 (Secrets) Store/Version,
 (Deployment) Ignore/Hide,
 (Secureness) Private Repository,
 (VCS Feature) History Sanitize,
 (Deployment) Store/Version,
 (Configuration File) Ignore/Hide,
 (Pre-open Source) Cross-check,
 (Client-Side Application) Exploitability,
 (Configuration File) Store/Version,
 (VCS Feature) Line Level Security,
 (Client-Side Application) Hide,
 (Secret) Ignore/Hide,
 (Configuration File) Accessibility,
 (Deployment) Dot File,
 (VCS Feature) Ignore Already Committed,
 (Configuration File) Exploitability,
 (Others) Decision,
 (Secureness) Unpushed Branch,
 (VCS Feature) Encrypt File
 \\
 \hline
\end{tabularx}
\end{table*}

\subsubsection{\uline{Answer to RQ1.3: Which questions are the most popular among developers related to checked-in secrets?}}

The popularity of each question category is presented in the ``PQ'' column of Table \ref{question-count-pct}. In our study, the popularity score varies between 0.005 and 0.030. For example, a question with Score 0 and View Count 17 has a PQ score of 0.005, whereas a question with Score 12 and View Count 17847 has a PQ score of 0.030. The top three most popular question categories are ``(Client-Side Application) Store'', ``(Secrets) Store/Version'' and ``(Client-Side Application) Hide''. In Table \ref{question-cat-hierarchy}, we also provide the question categories in descending order, sorted by PQ and UNC(\%). Further observations are aided by the ranking of the 27 question categories:

\begin{itemize}
    \item ``(Client-Side Application) Store'' and ``(Client-Side Application) Hide'' rank first and third based on the popularity score (PQ) and have a UNC score of 60.7\% and 35.7\%, respectively. The observation indicates that the questions related to storing and hiding secrets in client-side applications are most popular among developers but do not receive satisfactory answers. Therefore, future research is needed on the client-side frameworks for securely managing secrets.
    
    \item ``(Secrets) Store/Version'' ranks second based on the popularity score (PQ) and has a UNC score of 47.1\%. Our observation indicates that developers are showing more interest in the question of securely storing secrets for different technology frameworks such as ASP.NET, Ruby on Rails and Python. But, developers could not implement properly because of lacking proper documentation.
    
    \item ``(Secrets) Distribute'' and ``(Deployment) Improper Configuration'' question categories rank fourth and sixth for unsatisfactory answers, respectively. However, these question categories rank $22^{nd}$ and $19^{th}$ based on popularity score. Though the popularity score is low, developers are not receiving satisfactory answers for distributing secrets and fixing improper configuration errors during deployment. Therefore, future research can address secure secret distribution, and respective technology providers can provide proper documentation to fix improper configuration errors during deployment.
    
    \item We observe developers searching for VCS features to ignore the tracking of already-committed files to avoid local changes being accidentally committed in the VCS repository. An option exists to delete the file from remote repository and then ignore the file by placing the file name in the .gitignore file. However, developers do not want to delete and want a copy of the file in the remote repository, which VCS does not support~\cite{gitignore-atlassian}. Developers are also looking for line-level restrictions in VCS to hide secrets in particular lines of the source code. Though VCS has a feature called git smudge-clean~\cite{git-smudge-clean} which can be used to replace a secret with a dummy value during commits, developers face difficulties in implementing the process. Despite ``(VCS Feature) Ignore Already Committed'' and ``(VCS Feature) Line Level Security'' ranking $25^{th}$ and $24^{th}$, respectively, based on popularity score, the two question categories consist of 25 questions where developers are seeking the new VCS feature.
\end{itemize}

\subsubsection{\uline{Answer to RQ1.4: How do question categories related to checked-in secrets trend over time?}}

\begin{figure*}[!t]
    \includegraphics[width=\textwidth]{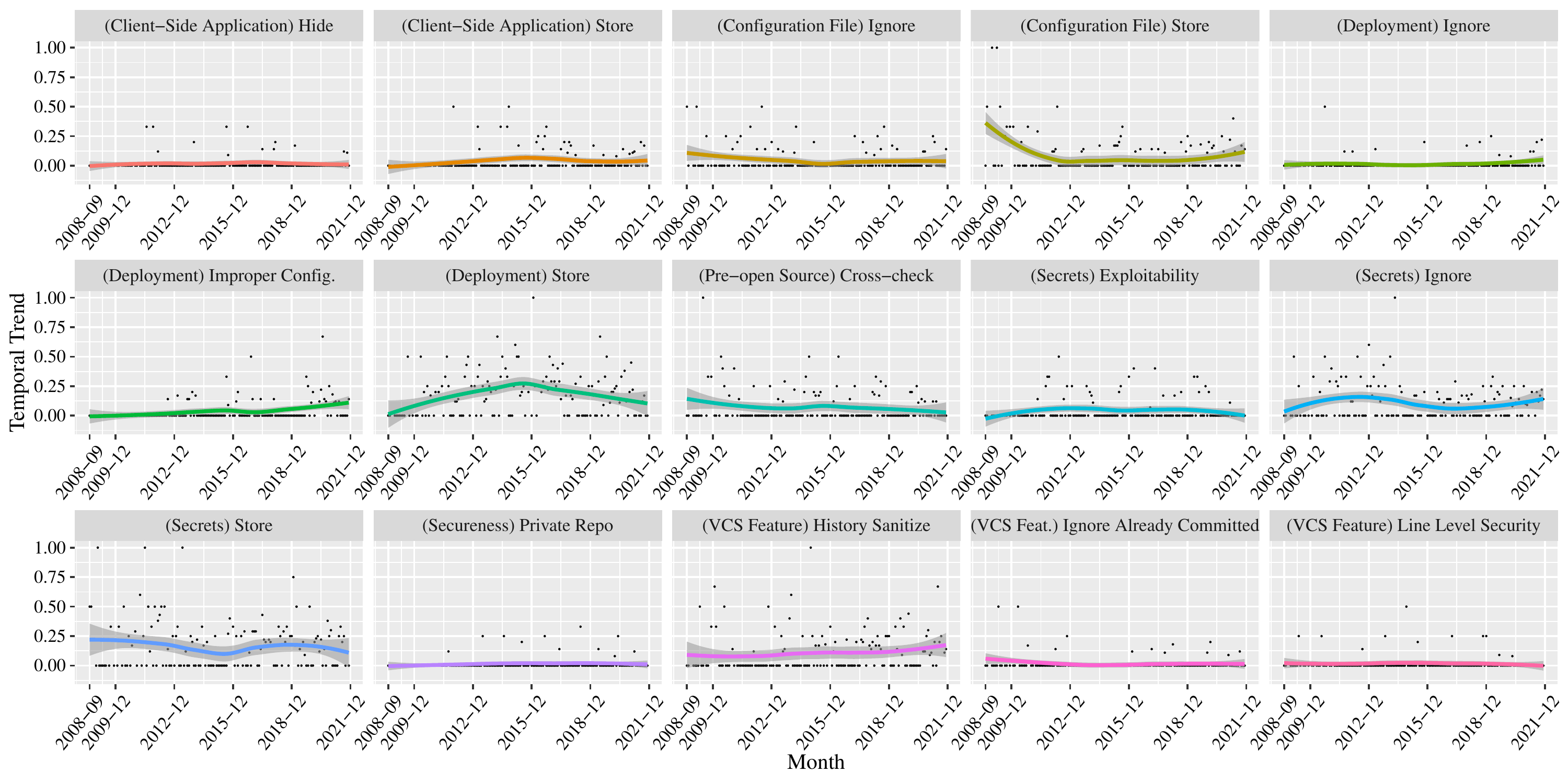}
    \caption{Temporal Trend of each identified question category. The month of the x-axis is shown in three-year interval. The zero value of Temporal Trend indicates no question is posted on the specific month for a category.}
    \label{fig:temporal_trend_question_category}  
\end{figure*}

Figure \ref{fig:temporal_trend_question_category} depicts the temporal trend of 15 question categories that have at least 10 questions. For each category, the figure provides a scatter plot with a smoothing plot with the trends highlighted. We can understand whether the trend of each question category is increasing, decreasing, or consistent from the ``Cox Stuart'', ``p-value'' and ``Trend'' columns of Table \ref{question-count-pct}. Table \ref{question-count-pct} highlights the question categories with a p-value less than 0.05 in grey.

From Table \ref{question-count-pct}, we observe an increasing trend in four question categories. While only four question categories showed increases, the trend is across 13 years of the data. We also observe that developers are posting more questions in ``(Secrets) Store/Version'', ``(VCS Feature) History Sanitize'', ``(Deployment) Improper Configuration'', and ``(Client-Side Application) Store'' categories, but their questions are not well-answered. 
The four question categories have a UNC score of more than 45\%, and three out of four question categories are also in the top six categories based on the popularity score (PQ). The increasing trend of these four question categories substantiates the absence of proper documentation on managing secrets during the deployment and the need for future research on client-side frameworks. In addition, the increasing trend also substantiates the need to improve existing VCS history sanitizing tools to make integration easier for developers.

\subsection{Answer to RQ2: What solutions do developers get for mitigating checked-in secrets?} 
\label{answer-categories}
We identify 13 answer categories from our analysis, which we present below based on the descending order of the number of questions in which StackExchange users suggest the specific answer category. For example, 179 answers to the 779 questions suggest the `A1: Move Secrets out of Source Code/Version Control and Use Template Config File' category. We do not declare all the answer categories as best practices. Indeed, below we highlight the shortcomings of these answer categories as appropriate.

\textbf{A1: Move Secrets out of Source Code/Version Control and Use Template Config File (179):} Developers may put secrets, such as database credentials, in a file where the code for database functionalities are present. As a result, developers face challenges in hiding the credentials from VCS repositories. In such cases, developers are suggested to move the secrets to a config file. Then, the config file with original secrets should be ignored from the VCS repository, and a template config file should be committed to the repository. Template config files, such as database.sample.yml file of Ruby on Rails, contain the minimum configurations with dummy secrets to avoid build failure. Developers will replace the dummy secrets in their development environment. Furthermore, a .gitignore file should be included with all repositories to ignore the secrets-containing files. GitHub has published a collection of .gitignore templates~\cite{GitHub-gitignore-template} for different technologies.    

\textbf{A2: Secret Management in Deployment (78):} We observe that developers mostly face challenges storing or versioning secrets for multiple environments during deployment. Configuration management systems, such as Ansible-Vault~\cite{ansible-vault} and Chef-Vault~\cite{chef-vault}, provide support for secret management. Developers are advised to use deployment variables, such as Heroku Config Vars~\cite{heroku-config-vars}, which create environment variables for respective environments. Developers are also suggested to keep the dot files such as .git and .hg files out of the root directory during deployment to avoid exposing secrets.

\textbf{A3: Use Local Environment Variables (56):} An environment variable is a dynamic object which is set outside of the application and used to avoid the storage of secrets in code or local config files. Developers are suggested to use environment variables to load the secrets at runtime. The benefits of using environment variables are switching secrets between deployed versions without modifying any code and making it less likely that secrets get checked into the repository. However, environment variables can leak secrets as they are passed down to child processes, which allows for unintended access~\cite{env-bad}.

\textbf{A4: Rewrite VCS History (48):} Secrets will not be removed entirely by removing in another commit as secrets will remain in the VCS history. Developers suggest removing secrets using git-filter-repo~\cite{git-filter-repo}, git-filter-branch~\cite{git-filter-branch}, and BFG repo cleaner~\cite{bfg-repo-cleaner}. Though official GitHub documentation~\cite{GitHub-rewrite-doc} suggests using BFG repo cleaner instead of git-filter-repo and git-filter-branch, we have seen Stack Exchange users mostly suggest using the latter. GitHub has also suggested contacting them with the repository name to clear the secrets from their cache and advised to tell the project collaborators to do git rebase instead of git merge~\cite{GitHub-rewrite-doc} though no Stack Exchange users' solutions suggested these actions.  

\textbf{A5: Store Encrypted/Obfuscated Secrets (39):} Storing secrets as encrypted, encoded, or obfuscated is one of the solutions suggested by Stack Exchange users. Different encryption algorithms, such as AES and RSA, are suggested. In some cases, developers are suggested to encode secrets using Base64 encoding in Android applications. Another suggestion is to split the secrets into multiple parts and keep them in the source code. The number of parts should be high, so the attacker will have to check for more than a billion permutations. Tools such as git-secret~\cite{git-secret} and git-crypt~\cite{git-crypt}, are available for encrypting secrets-containing files. The disadvantage of encryption is to deal with the encryption keys securely.

\textbf{A6: Use of External Secret Management Service (26):} Developers are recommended to implement external secret management services, such as HashiCorp Vault~\cite{hashicorp-vault} and AWS KMS~\cite{aws-kms}. These hardware security modules can safely store secrets with tightly-controlled access. However, because they are challenging to set up and maintain, these solutions may be unsuitable in some situations. In addition, they need a significant investment of time and money.

\textbf{A7: Load Externally and Use Secondary Private Repository (23):} Since developers want to avoid committing secrets into VCS that are needed for the application's functioning, developers are advised to load secrets externally using AWS S3 or a secondary private repository. Since AWS S3 needs access keys to retrieve stored files, the same problem of storing the access keys may occur. A secondary private repository can be used to store secrets and loaded dynamically using git submodules~\cite{git-submodule}. However, private repositories are not free from exploitation by attackers~\cite{private-repo-leak}.

\textbf{A8: Revocation and Rotation (16):} The first step to stop secrets sprawl is to revoke the secrets immediately. One developer suggested: \textit{``The important bit: Consider your credentials compromised. Change them. No matter what you do at this point, they are no longer secure''}~\cite{figshare-links}. A good practice is to rotate the secrets periodically. Short-lived secrets prevent previously-undetected data breaches from posing a threat, as access will be cut off even if the breach is not identified.  

\textbf{A9: Server-Side Implementation (16):} To avoid keeping secrets in client-side applications for fetching data from web services, developers are recommended to implement web service functionality on the server side. Then, the server will use the secrets and fetch data for the client side, thus removing the necessity to keep secrets in client-side applications.

\textbf{A10: VCS Feature (Git Hooks and Flags) (10):} To avoid secrets from pushing in VCS repositories, developers are suggested to implement git hooks~\cite{git-hook} and git flags~\cite{git-flag-assume-unchanged, git-flag-skip-worktree}. The pre-commit and post-commit hooks can be used to filter and smudge before commit or after pull, respectively~\cite{git-smudge-clean}. However, developers are warned as implementing git hooks properly is difficult. Developers are also suggested to use the git flags such as --skip-worktree~\cite{git-flag-skip-worktree} and --assume-unchanged~\cite{git-flag-assume-unchanged} to prevent changes from being committed to existing files.

\textbf{A11: Add Files to the Staging Area Explicitly (3):} A simple strategy to avoid exposing secrets accidentally is to add files explicitly in the VCS staging area. Developers are suggested to avoid using wildcards (git add -A or git add *) for adding files, thus having complete control and visibility over what files are committed.

\textbf{A12: Restrict API Access and Permissions (3):} Since attackers frequently use secrets within their scope, detecting when they are doing so maliciously might be challenging. However, damage and lateral movement can be limited by restricting access and permissions of the secrets. For example, GitHub IP white-listing~\cite{GitHub-white-list} can be employed to prevent any untrusted sources from accessing the GitHub repositories.

\textbf{A13: VCS Scan Tools (1):} Developers are advised to run VCS scan tools, such as TruffleHog~\cite{trufflehog} and Gitrob~\cite{gitrob}, before any commit or in an existing repository to find out the presence of secrets. The tools can find secrets buried in histories that manual searches and reviews will miss. However, tools may return a significant number of false positives~\cite{rahman2022secret}.

The mapping of answers to each question category can be found online~\cite{figshare-links-category-mapping}. We observe that the same answer category has been mentioned to mitigate challenges of multiple question categories. For example, `A1: Move Secrets out of Source Code/Version Control and Use Template Config File', `A3: Use Local Environment Variables' and `A2: Secret Management in Deployment' have been mentioned as part of a solution in 20, 12, and 10 out of 27 question categories, respectively.

\section{Discussion and Recommendations} \label{Discussion}
Below we discuss our findings and make recommendations.  In our discussion, we trace the questions and answers by their identifiers assigned in Table \ref{question-desc-example} and Section \ref{answer-categories}, respectively. 

\textbf{Tool enhancement.} We find that developers face difficulty with properly sanitizing VCS history (Q10). Developers commonly use git-filter-branch~\cite{git-filter-branch} and git-filter-repo~\cite{git-filter-repo} to sanitize VCS history. However, both the tools have safety and usability issues which can easily corrupt the repository's history~\cite{history-sanitize-safety}. For example, these tools can easily mix up the old and new history of the repository. In addition, coming up with the correct shell script is difficult as developers find out if the sanitizing code script is right or wrong by trying the script out. Even worse, broken filters often result in silent incorrect rewrites without proper output. Even if the developers sanitize the VCS history properly using the tools, the tools can not clear the cache in the respective version control systems, such as GitHub, as the sensitive information can appear again from the cache, according to GitHub's official documentation[3]. As of now, clearing from the cache is a manual process that can be automated.

In addition, we observe that developers are suggested to use VCS scan tools (A13) to avoid accidentally committing secrets, but developers seem to bypass scan tool warnings due to high false positives~\cite{rahman2022secret}. There are currently many open-source and proprietary VCS scan tools~\cite{list-secret-detection-tools}, but developers find it challenging to choose one tool out of many. Researchers and tool developers can work on comparing the effectiveness and efficiency of the VCS scan tools and improving the tools by reducing false positive warnings.

We also found that developers want new VCS features, such as line-level security, where developers can quickly point to the specific lines to which they want to restrict visibility in the VCS (Q12). In addition, we found that developers want to ignore local changes of already-committed files from VCS tracking without removing the file from the repository (Q11). Though Stack Exchange users suggested using --assume-unchanged~\cite{git-flag-assume-unchanged} and --skip-worktree~\cite{git-flag-skip-worktree} flags to ignore local changes of already-committed files from VCS tracking (A10), the official Git documentation suggests these flags not be used~\cite{ignore-already-committed}.

\begin{tcolorbox}[colback=black!6!white,colframe=black!6!white,boxrule=0.0mm, boxsep=0.1mm]
\textbf{Recommendation 1:} We recommend improving the existing tools, such as making the integration of VCS history sanitizing tools easy for the developers and reducing VCS scan tool false positives. We also recommend developing new tools for line-level security and ignoring local changes of already-committed files.
\end{tcolorbox}

\textbf{Documentation.} We find that developers face challenges in securely managing secrets while developing with different technologies due to the absence of proper documentation (Q1, Q6-Q8). For example, Foursquare API documentation~\cite{foursquare-api-doc} suggests developers use a client secret in userless or server-side authentication. However, a developer did not understand the documentation and asked in Stack Exchange whether the secret could be used in the client-side authentication~\cite{so-foursquare-api}. Developers also seem to query to understand the safest approach when multiple approaches are suggested in the same documentation~\cite{so-safest-approach}. For example, ASP.NET Core documentation suggests using local environment variables and secret manager tools to store secrets securely but does not specify which one will be the safest approach in specific use cases~\cite{asp-safest-approach}. However, we agree that no solution will be perfectly secure, but the documentation should be clear and detailed so that developers understand which use cases are appropriate for each approach. Furthermore, we observe that developers want reference links on how to implement a specific approach suggested in the documentation. For example, Google API provides documentation of the best practices for securely using API keys~\cite{google-api-external-link}. However, developers could not figure out how to implement these suggestions as reference links to the specific suggestions are not given~\cite{so-external-link}. We also observe that documentation does not explicitly mention whether the particular suggestion, such as setting up continuous deployment in Azure Function, is for the development or production environment~\cite{azure-production}. As a result, developers may implement a suggestion in the production environment that was intended for use in the development environment~\cite{so-production}, thus exposing secrets to the attackers. 

\begin{tcolorbox}[colback=black!6!white,colframe=black!6!white,boxrule=0.0mm, boxsep=0.1mm]
\textbf{Recommendation 2:} We recommend that each technology improve the technical documentation for managing secrets by i) clearly explaining the suggested approach's use cases and restrictions; ii) mentioning which approach will be safest for specific use cases when multiple approaches are suggested; iii) providing reference links to implement the suggested approaches; and iv) explicitly mentioning whether the particular approach is for development, production, or both environments. 
\end{tcolorbox}

\textbf{Client-side applications.}
Often, developers architect applications with only a client-side implementation and only later realize they must securely embed a secret in the code they distribute. As a result, questions about client-side secret storage (Q20), were the most popular among all topics we studied, as seen in Table \ref{question-count-pct}. One solution is for the developer to operate an API for their app that wraps the third-party API and keeps the secret server-side. Instead, novice developers embed third-party API calls in the client because it seems easier, cheaper (no infrastructure costs), and functions as expected. Unfortunately, secrets in the client-side application can not be protected against even a basic adversary with access to a debugger or decompiler. Inspired by popular DRM schemes such as Apple's FairPlay Streaming~\cite{AppleFairPlay}, we posit that privileged system elements, such as virtual machines, runtimes, browsers, or kernels could provide an interface for secure secret management.

\begin{tcolorbox}[colback=black!6!white,colframe=black!6!white,boxrule=0.0mm, boxsep=0.1mm]
\textbf{Recommendation 3:} 
We recommend that kernels and privileged runtimes develop frameworks to provide secure secret management for client-only applications.
\end{tcolorbox}

\textbf{Guidelines.} From the identified challenges in Table \ref{question-desc-example}, we observe that developers have a knowledge gap about whether a secret is exploitable or not (Q3), why they should keep secrets out of VCS (Q26), and what to do if they find secrets in the source code (Q27). We also found that some solutions are insecure for managing secrets by analyzing the solutions posed by Stack Exchange users. For example, storing secrets as Base64 encoded in the source code can be exploitable as secrets can be decoded easily (A5). Furthermore, storing secrets in a private repository is not a safe approach (A7) as private repositories are not free from exploitation by attackers or insider threats~\cite{private-repo-leak, nissan-leak}. Therefore, a guideline to train developers on securely managing secrets can eliminate the knowledge gap, and developers can make correct decisions during development. The National Institute of Standards and Technology (NIST)~\cite{nist} provides a framework SP 800-218~\cite{sp-800-218} to mitigate the risk of software vulnerabilities but does not have practices specific to securely managing secrets.

\begin{tcolorbox}[colback=black!6!white,colframe=black!6!white,boxrule=0.0mm, boxsep=0.1mm]
\textbf{Recommendation 4:} We recommend that NIST update the SP 800-218 framework by including practices specific to securely managing secrets to train developers.
\end{tcolorbox}

\section{Ethics} \label{Ethics}
The contents of all the Stack Exchange sites are under Creative Commons (CC BY-SA 3.0) license~\cite{cc-by-sa} with the following requirements: ``You are free to: \textit{Share} - copy and redistribute the material in any medium or format, \textit{Adapt} - remix, transform, and build upon the material for any purpose, even commercially''~\cite{cc-by-sa}. Stack Exchange inspires academics to utilize the data in research articles~\cite{jeff-academic} and requires researchers to give attribution to posts using a direct link~\cite{jeff-attibution}. As a result, we include hyperlinks to connect our quotes to the original posts, which are available online~\cite{figshare-links}.

\section{Threats to Validity} \label{ThreatToValidity}
In this section, we discuss the limitations of our paper. 

\textit{\uline{Q\&A Site Selection}}: We did not collect questions from other Q\&A sites, such as CodeProject \cite{codeproject} and Coderanch \cite{coderanch}. We accounted for this limitation by considering three Q\&A sites of Stack Exchange instead of only using Stack Overflow. 

\textit{\uline{Manual Analysis Bias}}: Caused by multiple interpretations and oversight, the manual analysis may induce bias. For example, the identified question and answer categories are susceptible to bias. We mitigated this bias by cross-checking the obtained question and answer categories and adding question and answer categories that both participants agreed on.

\textit{\uline{Closed Questions}}: The nature of inquiries about checked-in secrets in software artifacts may be broad, and Stack Exchange moderators do not like such questions. As a result, the moderators may decide to close some of the important questions. However, only 52 questions were flagged as closed, accounting for less than 7\% of the 779 questions in our dataset. We also observed that the closed questions had a high View Count (as high as 49471) and high Score (as high as 126) \cite{is-ex-pq}. As a result, we claim that the closed questions of our dataset have remained active after being closed, proving the significance of the topics under discussion. 

\textit{\uline{Popularity Metric}}: We measured the popularity metric of a question by taking the question's View Count and Score values into account. On the other hand, this metric may be biased because it ignores the time span of the views. Therefore, a new question with a low View Count and Score value may be regarded as unpopular. Also, Stack Exchange does not provide the temporal View Count of a question. As a result, a significant percentage of the View Count may accumulate when the question is initially posted or may have recently increased. Unfortunately, we have not yet arrived at a suitable treatment for this threat.

\textit{\uline{Counting Questions}}: We counted questions of a category posted by developers over time to find if a particular question category trends. There can be questions in that specific category that have been answered before, but developers are still posting new questions. It implies that the particular category continues to be a problem despite the ongoing effort. We agree that there can be a trend of decreasing questions of a category, but the problem may not be solved till today. However, we are not claiming those categories as of less importance. Instead, we are highlighting the recent ongoing problematic topics to the research community so that researchers can prioritize the challenges and work on resolving them.

\textit{\uline{Accepted Answer}}: We termed a question lacking an accepted answer as a \textit{question with unsatisfactory answer}. However, a developer who posted the question may be satisfied with the suggested solution posted by Stack Exchange users. Nevertheless, the developer may forget or not know how to mark the suggested solution as accepted in Stack Exchange. Unfortunately, we have not yet arrived at a suitable treatment for this threat.

\section{Related Work} \label{RelatedWork}
Prior work has found that root causes for widespread secret leakage were insecure practices, such as embedding hard-coded credentials~\cite{MedicalDataLeak,GithubLeaks}, organizational issues influencing software security vulnerabilities~\cite{assal2019think,xie2011programmers,nadi2016jumping}, and compromising security for functionality when managing software dependencies~\cite{pashchenko2020qualitative}. Researchers have looked into instances of such insecure developer practices within open-source projects~\cite{meli2019bad,7180102,rahman2019share,saha2020secrets,ding2020sniffing}. Researchers have discovered hard-coded secrets as a prevalent practice, resulting in thousands of repositories on open-source coding platforms, such as GitHub and Openstack, leaking hard-coded secrets~\cite{meli2019bad,rahman2019share, rahman2021different}. Within IaC scripts, Rahman et al.~\cite{rahman2019seven} looked for security smells, which are repeating coding patterns indicating a security flaw. They found 21,201 occurrences of seven security smells within 15,232 IaC scripts, and hard-coded credential is the most occurring smell with 1326 occurrences.

To understand more clearly the challenges that developers face, researchers have performed qualitative research into investigating what questions developers are asking on Stack Overflow (SO)~\cite{bajaj2014mining,rahman2019snakes,barua2014developers,rahman2018questions, TAHIR2020106333} as developers constantly search in SO for guidance on solving a challenge during development. Tahir et al.~\cite{TAHIR2020106333} looked through 4000 posts from three Stack Exchange sites to see what developers were discussing about code smells and anti-patterns. They observed that developers frequently post questions on Stack Exchange to check the presence of smell in their code, effectively using Q\&A sites as an informal code smell and anti-pattern detector.

We take motivation from the above studies and concentrate our research efforts on finding difficulties faced by developers for checked-in secrets in software artifacts. We also determine the solutions proposed by other developers to alleviate a specific challenge.

\section{Conclusion} \label{Conclusion}
Software relies heavily on the use of secrets for authentication and authorization, and the exposure of secrets is increasing each day. By analyzing the questions developers ask, we can understand the challenges developers face regarding checked-in secrets. In our empirical study, we studied 779 questions posted on Stack Exchange to investigate the challenges faced by developers and the corresponding solutions posed by others to mitigate the challenges. We identified 27 challenges and 13 solutions. The four most common challenges, in ranked order, are: (i) store/version of secrets during deployment (Q6); (ii) store/version of secrets in source code (Q1);  (iii) ignore/hide of secrets in source code (Q2); and (iv) sanitize VCS history (Q10). The three most common solutions, in ranked order, are: (i) move secrets out of source code/version control and use template config file (A1); (ii) secret management in deployment (A2); and (iii) use local environment variables (A3). In addition, we observe that the same solution has been mentioned to mitigate multiple challenges. We also observe an increasing trend in questions lacking accepted answers. Our findings will benefit researchers and tool developers who can investigate how the secret management process can be enhanced to facilitate secure development.


\section*{Acknowledgment}

This work was supported by National Science Foundation 2055554 grant.  The authors would also like to thank the North Carolina State University Realsearch research group for their valuable input on this paper.

%for later:  grant number 2055554.



\bibliographystyle{IEEEtran}
%\bibliographystyle{plain}
\bibliography{bibliography}

%\bibliographystyle{plain}
%\bibliography{bibliography}

% \section*{Appendix}
% \input{Tables/QuestionCountPerYear}
% \input{Tables/tagskeywords}

\end{document}